\documentclass{article}

     \PassOptionsToPackage{numbers, compress}{natbib}
\usepackage[dblblindworkshop, final]{neurips_2025}

\workshoptitle{Machine Learning and the Physical Sciences}

\usepackage[utf8]{inputenc} % allow utf-8 input
\usepackage[T1]{fontenc}    % use 8-bit T1 fonts
\usepackage{hyperref}       % hyperlinks
\usepackage{url}            % simple URL typesetting
\usepackage{booktabs}       % professional-quality tables
\usepackage{amsfonts}       % blackboard math symbols
\usepackage{nicefrac}       % compact symbols for 1/2, etc.
\usepackage{microtype}      % microtypography
\usepackage{xcolor}         % colors
\usepackage[pdftex]{graphicx}
\usepackage{placeins}  
\usepackage{amsmath}
\usepackage{array}
\usepackage{comment}

\title{Differentiable Physics-Neural Models enable Learning of Non-Markovian Closures for Accelerated Coarse-Grained Physics Simulations}

\author{%
  Tingkai Xue \\
  Department of Mechanical Engineering\\
  National University of Singapore\\
  Singapore 119007, Singapore \\
  \texttt{xue.tingkai@u.nus.edu} \\
  \And
  Chinchun Ooi \\
  Institute of High Performance Computing\\
  Agency for Science, Technology and Research\\
  Singapore 138632, Singapore \\
  \texttt{ooicc@cfar.a-star.edu.sg} \\
  \And
  Zhengwei Ge \\
  Institute of High Performance Computing\\
  Agency for Science, Technology and Research\\
  Singapore 138632, Singapore \\
  \texttt{gezw@ihpc.a-star.edu.sg} \\
  \And
  Fong Yew Leong \\
  Institute of High Performance Computing\\
  Agency for Science, Technology and Research\\
  Singapore 138632, Singapore \\
  \texttt{leongfy@ihpc.a-star.edu.sg} \\
  \And
  Hongying Li \\
  Dept of Mechanical and Aerospace Engineering\\
  Nanyang Technological University\\
  Singapore 639798, Singapore \\
  \texttt{hongying.li@ntu.edu.sg} \\
  \And
  Chang Wei Kang \\
  Institute of High Performance Computing\\
  Agency for Science, Technology and Research\\
  Singapore 138632, Singapore \\
  \texttt{kangcw@ihpc.a-star.edu.sg} \\
}

\begin{document}

\maketitle

\begin{abstract}
Numerical simulations provide key insights into many physical, real-world problems. However, while these simulations are solved on a full 3D domain, most analysis only require a reduced set of metrics (e.g. plane-level concentrations). This work presents a hybrid physics-neural model that predicts scalar transport in a complex domain orders of magnitude faster than the 3D simulation (from hours to less than 1 min). This end-to-end differentiable framework jointly learns the physical model parameterization (i.e. orthotropic diffusivity) and a non-Markovian neural closure model to capture unresolved, `coarse-grained' effects, thereby enabling stable, long time horizon rollouts. This proposed model is data-efficient (learning with 26 training data), and can be flexibly extended to an out-of-distribution scenario (with a moving source), achieving a Spearman correlation coefficient of 0.96 at the final simulation time. Overall results show that this differentiable physics-neural framework enables fast, accurate, and generalizable coarse-grained surrogates for physical phenomena.
\end{abstract}

\section{Introduction}
Numerical simulations, e.g. computational fluid dynamics (CFD), are a cornerstone for understanding complex physical phenomena, but their high computational cost can be extremely restrictive. Coarse-graining of simulations is a widely-used alternative to reduce computational cost, especially in multi-scale systems, whereby the dimensionality of the system is reduced through model parameterizations of unresolved fine-scale dynamics~\cite{parish2017non}. According to the Mori–Zwanzig formalism, this coarse-graining requires memory-dependent closure terms to better account for unresolved interactions across scales. Architectures such as Long Short-Term Memory (LSTM) model were used to predict temporal dynamics in a reduced order basis \cite{cnn_ae_lstm_turbulent, temporal_dynamics}, but accurate, generalizable prediction remains difficult. Recent work in scientific machine learning has also shown the promise of physics-informed surrogates~\cite{physics_informed_ml, sanderse2024scientific}, and neural closures have been proposed for modelling of dynamical systems, including subgrid closures for turbulent flows, and neural closures for incomplete multiphysics forecasting~\cite{doi:10.1073/pnas.2101784118, Melchers_2022,generalized_neural_closure, maulik2019subgrid, pan2018data}. However, these efforts rely on purely Markovian closure terms, ignore history-dependent effects, and/or treat physical parameters as fixed rather than trainable quantities. %While recent advances in scientific machine learning have shown promise, most approaches focus on a scalar performance metric (e.g. drag coefficient), or require large amounts of training data (e.g. $> O(10^3)$, complicating their real-world applicability. 

Inspired by recent innovations whereby Physics guides ML, we propose a hybrid physics–neural surrogate model that provides orders-of-magnitude speed-up over conventional numerical simulations through an analogous coarse-grained-inspired approach for physical consistency. This end-to-end differentiable framework explicitly learns only the non-Markovian closure terms, which potentially explains the model's predictive and extrapolative ability even with a small training dataset ($N_{training} = 26$ 3D full-field simulations). Critically, the proposed model incorporates joint inference of improved physical parameterizations for encoding the physical environment (e.g. diffusivity), a critical step which was found to improve model performance.

The proposed methodology is applied to the transport of a released scalar (e.g. pollutant, airborne virus) in a complex real-world indoor environment. While the full 3D domain greatly impacts the physics, the scalar concentration at human height (1.5 m above ground level) is typically the key metric of interest. Hence, the `coarse-grained physics' allows acceleration through two key processes without affecting its practical utility: 1) Reduction of problem from the full 3D domain to a 2D plane of interest in the domain; 2) Model-learned implicit parameterization of domain geometry (e.g. obstructions like pillars, and non-linear effects from boundary conditions like the air-conditioners). 

For effective long-term simulation, the unresolved physics (e.g. flow orthogonal to the modeled plane) has to be well-learned by the proposed physics-neural model~\cite{incompletephysics}. By combining a finite-volume solver (physics-based) with an LSTM-based neural closure model, we provide a principled approximation of the memory effects as required by the Mori–Zwanzig formalism~\cite{ruiz2024benefits}. Moreover, the differentiable physics-neural framework allows physical quantities (e.g. the orthotropic diffusivity) to be learned simultaneously, permitting more physical model parameterizations~\cite{thuerey2021pbdl,effective_diffusivity}. We further demonstrate potential for generalization beyond training distributions, including moving-source scenarios, while maintaining orders-of-magnitude reductions in runtime relative to the full 3D numerical simulation (from hours to $<$ 1 min).

\section{Problem formulation}
Scalar transport is widely studied in urban environments (e.g. pollutant/chemical dispersion, airborne infectious disease). This physics can be encapsulated by the convection-diffusion equation (Eq.~\ref{eq:convection_diffusion}): 

\begin{equation} \label{eq:convection_diffusion}
\frac{\partial c}{\partial t} =
\nabla \cdot \Big( D \nabla c - \vec{v} \, c \Big) + R
\end{equation}

where $c:\Omega\times[0,T]\to\mathbb{R}_{\ge 0}$ is the scalar quantity of interest in the spatio-temporal domain denoted by $\Omega \subset \mathbb{R}^3$ and $t \in [0,T]$. $D$ is the (potentially orthotropic) diffusivity tensor, $\vec{v}$ is the background velocity field (e.g. derived from CFD), and $R$ denotes sources and sinks in the domain. Initial conditions $c(\cdot,0)=c_0$ and boundary conditions are pre-specified (e.g. zero flux condition on walls). While $c(x,y,z, t)$ is a 3D scalar field, the proposed hybrid physics-neural model solves for $\bar{c}(x,y,t)=c(x,y,h,t)$ for a fixed height $h$ as denoted by $\Omega_h$ ($h = 1.5m$ in the rest of this work).

%\begin{equation}
%\label{eq:plane-def}
%\Omega_h \;:=\; \{(x,y)\in\mathbb{R}^2:\,(x,y,h)\in\Omega\}, 
%\qquad
%\bar c(x,y,t) \;:=\; c(x,y,h,t),
%\end{equation}
%for $(x,y)\in\Omega_h$ and $t\in[0,T]$.

Projecting Eq.~\ref{eq:convection_diffusion} onto the 2D plane yields an effective 2D transport equation:
\begin{equation}
\label{eq:2d-pde-closure}
\frac{\partial \bar c}{\partial t} = \nabla_{xy}\!\cdot\!\Big(\bar D\,\nabla_{xy}\bar c - \bar{\vec v}\,\bar c\Big)
\;+\; \bar R
\;+\; \underbrace{\mathcal{M}_{\theta}\!\big(\bar c_{0:t},\,\Phi\big)}_{\text{non-Markovian closure}},
\quad (x,y)\in\Omega_h,\ t\in(0,T],
\end{equation}
where $\bar D$, $\bar{\vec v}$, and $\bar R$ are effective (plane-level) quantities, and $\mathcal{M}_{\theta}$ is a \emph{history-dependent} closure term that accounts for unresolved effects (e.g., transport normal to the plane), which depends on the historical scalar field $\bar c_{0:t}$ and some parameters $\Phi$ (e.g. geometry encodings). 

A Markovian module restricts $\mathcal{M}_\theta$ to depend only on $\bar c(\cdot,t)$ and limits long time horizon accuracy. In Section \ref{results}, we demonstrate that the learning of a non-Markovian closure model and the joint learning of $\bar{D}$ both significantly improve the model's prediction ability, such that qualitatively consistent simulations of out-of-distribution scenarios (e.g. moving source scenarios) can be obtained.

\section{Method}
\subsection{Dataset}
High-fidelity simulation data is obtained from a commercial software (ANSYS Fluent). Multiple convection-diffusion simulations are performed in a 3D indoor environment (with features such as air-conditioning, exhaust fans, and pillars) using the steady-state velocity $\vec{v}$ obtained by solving the governing equations with the imposed boundary conditions (constant outlet velocity at air-cons and zero flux condition on walls) and a constant passive scalar source term $R$ corresponding to a $0.1 \times 0.1 \times 0.1 m^3$ volumetric source. The height of the entire 3D domain is 2.46m, which is sufficient for recirculation to occur and appear as non-Markovian effects that need to be captured in a reduced 2D model. The total simulated time is 600s (10min). A total of 53 full-field simulations, each with its own source ($R(x,y)$) were generated, and 26 were used for training. 2D plane-slices were extracted on a $49\times 26$ structured grid ($\Delta x=\Delta y= 0.5m$), where each grid cell's value is the surface average concentration. The velocity fields ($\vec{v}$) and turbulent diffusivity fields ($D$) from the numerical model are provided in Appendix \ref{train_physics} Figure \ref{fig:physical_params}. % at various flow times with irregular time intervals

\subsection{Hybrid physics-neural model}\label{hybrid_model}
\begin{figure}
    \centering
    \includegraphics[width=0.8\linewidth]{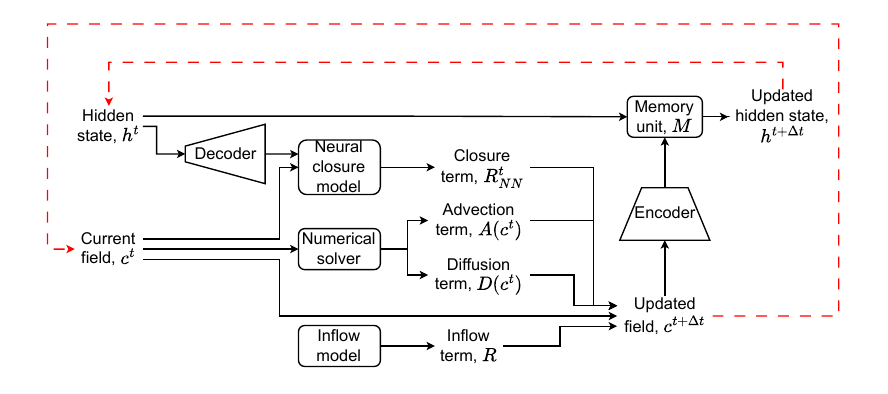}
    \caption{Schematic of time advancement scheme in the proposed hybrid physics-neural model. Dotted lines indicate the flow of hidden state and concentration field from each time step to the next.}
    \label{fig:hybrid_time_advancement}
\end{figure}

The hybrid physics-neural model has a solver-in-loop setup \cite{solver_in_loop} and is illustrated in Figure \ref{fig:hybrid_time_advancement}. Each time step involves a physics-based finite volume solver that computes the diffusion term $D(c^t;\Delta t)$ and advection term $A(c^t;\Delta t)$ from the current concentration field $c^t$, and a non-Markovian neural closure model for $R_{NN}^t$ that accounts for the unresolved physics in the form of an history-dependent change in concentration values. Given a specified release location, a corresponding source term $R$ is also provided. Details of each module are provided in \ref{nn_spec}. 

The finite volume method is used to solve the convection-diffusion equation. The explicit forward Euler method is used for time advancement while an upwind scheme is used for advection. An orthotropic diffusivity is assumed for computing the diffusion term. Numerical terms are computed using smaller time steps as required to avoid violation of the CFL condition. %This leads to a negligible increase in runtime., the diffusivity is assumed to be orthotropic, as diffusion between adjacent cells depends on the interface 

Two states are advanced in parallel - a hidden state $h^t$ containing the history of the system, as well as the passive scalar concentration field $c^t$. The hidden state is used to account for unresolved multi-scale effects over multiple time steps~\cite{Melchers_2022}. As the concentration ($c^{t+\Delta t}$) is being updated every time step as per Eq.~\ref{eq:update_field}, an LSTM module is simultaneously also being updated to yield a history-aware hidden state as per Eq.~\ref{eq:update_hidden}, which is a crucial input to the non-Markovian neural closure model. 
\begin{equation}\label{eq:update_field}
c^{t+\Delta t} = c^t + A(c^t,\Delta t) + D(c^t,\Delta t) + R \Delta t + R_{NN}(h^t;c^t)
\end{equation}
\begin{equation}\label{eq:update_hidden}
h^{t+\Delta t} =M(h^t,c^{t+\Delta t})
\end{equation}
% This constitutes one discrete constant time step $\Delta t$ which will be repeated for longer time intervals. This model is a discretized version of Augmented Neural ODE (ANODE) presented in \cite{anode}. ANODE is not used as it is found to be slower and adaptive time stepping may lead to numerical errors for the convection and diffusion terms.
%\subsubsection{Numerical methods for convection-diffusion}\label{numerical_methods}
%\subsubsection{Neural closure model} \label{neural_network_closure}
 
At each time, the hidden state is first decoded using transpose convolutional layers, which are then concatenated with the current concentration field. A convolutional neural network (CNN) is used to compute the closure term due to its suitability for grid-like data and ability to extract local spatial information at various scales \cite{NIPS2012_c399862d}. The closure term should also account for local effects, such as advection and diffusion normal to the plane. As such, local properties, i.e. velocity and learnt diffusivity fields, are also provided as inputs. The combination of the closure term, advection term, diffusion term, and source term gives the time-stepped concentration field, $c^{t+\Delta t}$, which is the input to the memory unit to get the time-stepped hidden state $h^{t+\Delta t}$. An LSTM model is used due to its well-established ability to model long-term dependencies \citep{lstm}. 

\subsection{Training}
% differentiable physics approach
The trainable parameters in this project are the neural network parameters as well as the orthotropic diffusivity values. The diffusivity values are trained as they should reflect unresolved internal features (e.g. pillars) and is in line with the concept of `coarse-grained' physics. 

There are two main difficulties. Firstly, due to the large number of steps in the prediction sequence, unrolling the whole sequence of data for backpropagation through time would require a lot of RAM. Secondly, two parallel but inter-dependent sequences ($c^t$ and $h^t$) are trained simultaneously. Therefore, the 600s of flow time is broken into 60s segments. Except for the first 60s (for which the LSTM states and concentration fields are initialized as zeros), subsequent segments start from the predicted hidden state and concentration field from the previous segment. Within each segment, the prediction at time steps for which a desired concentration field is available in the training dataset is used to compute the loss. Additionally, curriculum learning \cite{curriculum_learning} is employed to train the model with simulations of increasing duration. The model is trained for 500 epochs at each stage, using simulations of 60s, 120s, 300s, and 600s. More details on model training are in Appendix \ref{training_specification}. 

Additional ablation studies are conducted to compare (i) the hybrid-neural model vs a data-driven model, and (ii) the non-Markovian vs Markovian model. The standard deviation across triplicate runs with different seeds are also presented as error bars in all subsequent plots. The dataset and the training code are available on https://github.com/IHPC-Physics-AI/Closure-for-Convection-Diffusion. %Experimental replicability is tested by initializing the neural network parameters with different seed values, and the standard deviation across the triplicates is presented as error bars in the corresponding plots

\section{Results} \label{results}

\textbf{Hybrid physics-neural approach helps with data-efficient training and physical consistency.} As shown in Appendix \ref{baseline_performance} Fig.~\ref{fig:generalization}, the data loss decreases with training. After training with all 600s of flow time, both the train and test data loss decreased, especially for the longer time rollouts. The initial high loss may be due to numerical errors that led to large predicted concentrations at low concentration regions, thereby increasing the relative error. Fig.~\ref{fig:data_efficiency} further shows that the model is able to start learning and improving, even with just 8 data points (i.e. is data efficient).

Qualitatively, the results are consistent with the physical system, e.g. learnt diffusivity for regions with walls is generally lower, as exemplified by the region circled in red in Fig.~\ref{fig:qualitative}(d) which resembles the physical layout in Fig.~\ref{fig:qualitative}(a). A vertical line of low diffusivity is also circled in Fig.~\ref{fig:qualitative}(c) which is related to an "air curtain"-like effect caused by the two air-cons nearby. while the closure term in Fig.~\ref{fig:qualitative}(b) shows negative values at the air-con positions, which concurs with our understanding that air-cons lead to reduction in passive scalar concentration from the plane. %It can be observed that some aspect of the physical domain is learned. Appendix \ref{baseline_performance} 

A purely data-driven model is also trained by removing the numerical solver and inflow module, and only retaining the closure term, i.e. $c^{t+\Delta t}=c^t+R_{NN}^t\Delta t$. The resultant loss is significantly higher (Fig.~\ref{fig:original_v_datadriven}) and predictions are highly non-physical (Fig.~\ref{fig:predict_original_v_datadriven}). Results are in Appendix \ref{physics_informed_v_data_driven}. 

\textbf{Including memory in the model improves prediction.} A purely Markovian model is tested by removing the hidden state and memory unit, such that the closure term only depends on the current scalar field. As per Appendix \ref{nm_v_markov}, although the loss is comparable initially, the loss at later flow times is higher, as is consistent with the intuition that the memory effect is most obvious at longer time horizons. As observed from Figure \ref{fig:predict_original_v_markovian}, features are also significantly different for the predicted closure term, implying that memory is important for learning the non-Markovian effects.

\textbf{Joint learning of physical parameters improves performance.} We further compare the effect of joint learning of physical parameterizations such as diffusivity and velocity, and noticed that model performance is enhanced by incorporation of this component, as per Appendix \ref{train_physics}.

\textbf{The model is promising for out-of-distribution scenario.} The trained model is used to predict evolution with a moving source, unlike simulations in the training set with static release locations. The relative loss is much lower than a baseline method, where concentration is assumed homogeneous and equal to the average concentration throughout the whole simulation (Appendix \ref{moving_case_scenario} Fig.~\ref{fig:transient_compare}). The prediction is also more meaningful as the model's prediction shows similar features to the numerical simulation (Fig.~\ref{fig:transient_snapshots}). There is a strong correlation (Spearman correlation coefficient $=0.96$) between predicted concentration and actual concentration, even at the final time step (Fig.~\ref{fig:transient_correlation}). This means the model can be used to identify regions of higher concentration.

\section{Discussion and Limitations}\label{limitations}
We demonstrated that a hybrid physics-neural model can deliver fast, long time horizon predictions on a reduced domain (2D plane) while preserving physical consistency through a learned, non-Markovian neural closure model. The physics-neural design proved more data-efficient and physically consistent than its data-driven equivalent, and explicitly modeling memory improved accuracy versus Markovian baselines. Predictions retained meaningful structure and correlated strongly with ground truth, even on a moving source scenario (not in training data). Finally, the learned orthotropic diffusivity qualitatively aligns with aspects of the physical environment (e.g., walls/air-curtain–like regions), reinforcing the value of jointly learning model parameters with the non-Markovian closure.

However, some areas remain for further improvement. Early-time errors on the coarse-grained simulation remain large, which may have resulted from explicit, first-order updates. While it was hypothesized that the closure model may be able to compensate for these errors, the errors remain larger. Also, additional studies to evaluate the generalizability of this methodology (e.g., with different boundary conditions or other physics problems) would provide further insight into areas for improvement. Furthermore, the model's sensitivity to different parameters, such as LSTM size, memory horizon, or grid resolution (i.e. extent of coarse-graining), could all be studied to better understand the model's limitations and capacity for generalization. During the training of our model, a penalty (in the form of L2 regularization) is applied on the neural closure model to constrain the closure model and prevent over-fitting; however, the learned diffusivity and neural closure model remains highly under-constrained. Additional ablation studies may provide greater insight into their respective impact, and better identifiability of the contribution of each of the numerical and closure terms could inspire additional strategies to increase the data-efficiency and predictive performance of the proposed model. 

\medskip
\bibliographystyle{ieeetr}
\bibliography{aaai25}

%%%%%%%%%%%%%%%%%%%%%%%%%%%%%%%%%%%%%%%%%%%%%%%%%%%%%%%%%%%%
\newpage
\appendix
\section{Technical Appendices and Supplementary Material}
\subsection{Architecture of neural network modules}\label{nn_spec}
This section details the modules composing the model outlined in Fig.~\ref{fig:hybrid_time_advancement} and their respective neural architectures. In all modules, LeakyReLu with negative gradient of 0.01 is used as the non-linear activation. KS refers to the kernel size used.

Briefly, the inflow module in Table~\ref{tab:inflow_spec} infers boundary inflows from a user-defined source location. The encoder in Table~\ref{tab:encoder_spec} compresses the current concentration field into a compact latent state while the decoder in Table~\ref{tab:decoder_spec} upsamples the hidden state to the actual physical field resolution so it can be concatenated with the current field variables. Separately, the LSTM in Table~\ref{tab:memory_spec} ensures the long-time propagation of history across steps in the hidden state, a crucial feature for ensuring the non-Markovian neural closure model in Table~\ref{tab:closure_spec} has sufficient information for incorporating unresolved physics. 

\begin{table}[h]
    \centering
\caption{Inflow module architecture}
\label{tab:inflow_spec}
    \begin{tabular}{lcccc}\toprule
           Layer no.&Layer type&  Input shape&  Specifications&  Output shape\\\midrule
           1&Conv2D&  (26,49,7)&  KS=(3,3)&  (26,49,8)\\
           2&MaxPool2D&  (26,49,8)&  KS=(2,3), strides=(2,3)&  (13,16,8)\\ 
   3&Conv2D& (13,16,8)& KS=(3,3)& (13,16,16)\\
   4&MaxPool2D& (13,16,16)& KS=(2,3), strides=(2,3)& (6,5,16)\\
   5&Conv2D& (6,5,16)& KS=(3,3)& (6,5,32)\\
   6&MaxPool2D& (6,5,32)& KS=(2,2), strides=(2,2)& (3,2,32)\\
   7&Conv2D& (3,2,32)& KS=(3,3)& (3,2,64)\\
   8&MaxPool2D& (3,2,64)& KS=(3,2), strides=(3,2)& (1,1,64)\\ 
 9& Fully Connected& (1,1,64)& features=64&(1,1,64)\\
 10& Fully Connected& (1,1,64)& features=64&(1,1,64)\\
 11& Fully Connected& (1,1,64)& features=4&(1,1,4)\\ \bottomrule
    \end{tabular}
\end{table}

\begin{table}[h]
    \centering
\caption{Encoder module architecture}
\label{tab:encoder_spec}
    \begin{tabular}{lcccc}\toprule
           Layer no.&Layer type&  Input shape&  Specifications&  Output shape\\\midrule
           1&Conv2D&  (26,49,3)&  KS=(3,3)&  (26,49,16)\\
           2&Conv2D&  (26,49,16)&  KS=(3,3)&  (26,49,16)\\
   3&MaxPool2D& (26,49,16)& KS=(2,3), strides=(2,3)& (13,16,16)\\
   4&Conv2D& (13,16,16)& KS=(3,3)& (13,16,32)\\
   5&Conv2D& (13,16,32)& KS=(3,3)& (13,16,32)\\
   6&MaxPool2D& (13,16,32)& KS=(2,3), strides=(2,3)& (6,5,32)\\
   7&Conv2D& (6,5,32)& KS=(3,3)& (6,5,64)\\
   8&Conv2D& (6,5,64)& KS=(3,3)& (6,5,64)\\
 9& MaxPool2D& (6,5,64)& KS=(2,2), strides=(2,2)&(3,2,64)\\
 10& Conv2D& (3,2,64)& KS=(3,3)&(3,2,128)\\
 11& Conv2D& (3,2,128)& KS=(3,3)&(3,2,128)\\
 12& MaxPool2D& (3,2,128)& KS=(3,2), strides=(3,2)&(1,1,128)\\ \bottomrule
    \end{tabular}

\end{table}

\begin{table}[h]
    \centering
\caption{Decoder module architecture}
\label{tab:decoder_spec}
    \begin{tabular}{lcccc}\toprule
           Layer no.&Layer type&  Input shape&  Specifications&  Output shape\\\midrule
           1&ConvTranspose&  (1,1,512)&  KS=(3,2), strides=(3,2)&  (3,2,128)\\
           2&Conv2D&  (3,2,128)&  KS=(3,3)&  (3,2,128)\\ 
   3&ConvTranspose
& (3,2,128)& KS=(2,2), strides=(2,2)& (6,5,64)\\
   4&Conv2D& (6,5,64)& KS=(3,3)& (6,5,64)\\
   5&ConvTranspose
& (6,5,64)& KS=(2,3), strides=(2,3)& (13,16,32)\\
   6&Conv2D& (13,16,32)& KS=(3,3)& (13,16,32)\\
   7&ConvTranspose
& (13,16,32)& KS=(2,3), stride=(2,3)& (26,49,16)\\
   8&Conv2D& (26,49,16)& KS=(3,3)& (26,49,16)\\ \bottomrule
    \end{tabular}
    
\end{table}

\begin{table}[h]
    \centering
\caption{Memory unit architecture}
\label{tab:memory_spec}
    \begin{tabular}{lcccc}\toprule
           Layer no.&Layer type&  Input shape&  Specifications&  Output shape\\\midrule
           1&LSTMCell&  (1,1,128)&  (1,1,128)&  (1,1,128)
\\
           2&LSTMCell&  (1,1,128)&  (1,1,128)&  (1,1,128)
\\ 
   3&LSTMCell& (1,1,128)& (1,1,128)& (1,1,128)
\\
   4&LSTMCell& (1,1,128)& (1,1,128)& (1,1,128)\\ \bottomrule
    \end{tabular}
    
\end{table}

\begin{table}[h]
    \centering
\caption{Neural closure module architecture}
\label{tab:closure_spec}
    \begin{tabular}{lcccc}\toprule
           Layer no.&Layer type&  Input shape&  Specifications&  Output shape\\\midrule
           1&Conv2D&  (26,49,26)&  KS=(3,3)&  (26,49,26)\\
 2& Conv2D& (26,49,26)& KS=(3,3)&(26,49,26)\\
 3& Conv2D& (26,49,26)& KS=(3,3)&(26,49,1)\\ \bottomrule
    \end{tabular}

\end{table}
\FloatBarrier

\subsection{Training specifications} \label{training_specification}

\begin{figure}[h]
    \centering
    \includegraphics[width=0.9\linewidth]{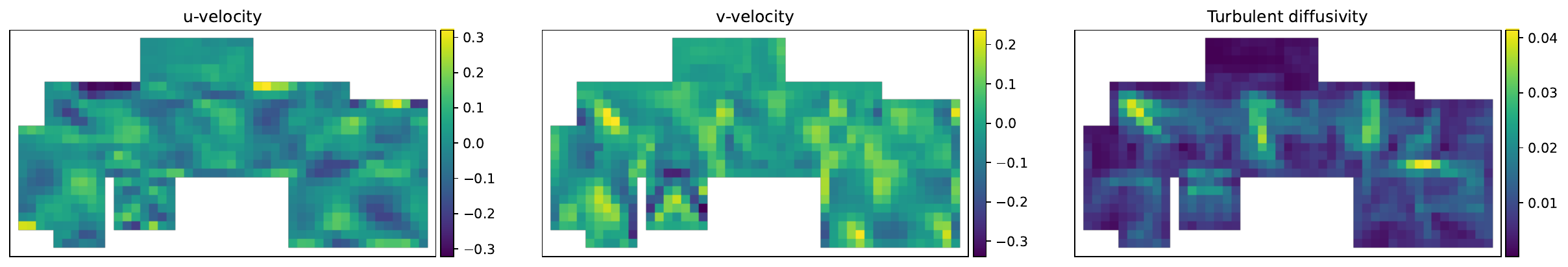}
    \caption{Velocity fields and turbulent diffusivity field (computed as $\nu_t/Sc_t$, where $\nu_t$ is the turbulent viscosity and $Sc_t=0.9$ is the turbulent Schmidt number) as extracted from the full 3D numerical simulation}
    \label{fig:physical_params}
\end{figure}

More details on the model training settings and hyperparameters are provided in this section. All models are implemented in JAX and trained on a single Nvidia RTX 3090 GPU card. The neural network weights are initialized randomly using Glorot uniform initializers, while the diffusivity values are initialized using the turbulent diffusivity values extracted from the 3D numerical simulation model as per Fig.~\ref{fig:physical_params} and trained jointly with the rest of the network. Optimization is performed using the Adam optimizer using a constant learning rate of $10^{-5}$. Each minibatch consists of 5 simulation segments of 60s flow time. 

The combined loss function is calculated as per Eq. \ref{eq:loss_equation}:
\begin{equation}\label{eq:loss_equation}
L=L_{data}+\lambda_1 |w|^2 +\lambda_2 \sum_{t=1}^{n_t-1} \sum_{i=0}^{n_x-1}\sum_{j=0}^{n_y-1}(u_{i,j}^t-u_{i,j}^{t-1})^2
\end{equation}

where the data loss $L_{data}$ is calculated as the pixel-wise square of the relative error of each cell:

\begin{equation}\label{eq:loss_equation_data}
L_{data}=\frac{1}{n_x n_y}\sum_{t=0}^{n_t-1}\sum_{i=0}^{n_x-1} \sum_{j=0}^{n_y-1}\left(\frac{u_{i,j}^t-\hat{u}_{i,j}^t}{\hat{u}_{i,j}^t+\varepsilon}\right)^2I(\text{$u_{i,j}^t$ is avaliable})
\end{equation}

Here, $n_t,n_x,n_y$ are the number of time steps and number of pixels in the horizontal and vertical directions respectively. $u$ is the predicted concentration while $\hat{u}$ is the ground truth concentration. $I(\cdot)$ is the indicator variable giving 1 if the statement is true and 0 otherwise. As the pixel-wise concentration may be zero, a small constant of $\varepsilon=10^{-9}$ is added in the denominator to prevent the loss term from diverging as the true value approaches zero. Additionally, L2 weight decay is applied to the neural closure model's weights $w$ with a constant factor of $\lambda_1=10^{-6}$. This is motivated by \cite{leads} where a penalty was imposed on the case-specific component to maximize the learning of the component that is shared across different cases, which in our case is the diffusivity value. 

It is also desirable that predictions not oscillate. Therefore, a regularization term is imposed in terms of the pixel-wise sum of squares of the change between consecutive predictions with a constant factor of $\lambda_2=10^4$. The magnitude of $\lambda_2$ is relatively higher as the values of the scalar concentrations in this problem are on the order of $10^{-5}$ to $10^{-6}$.

\FloatBarrier

\subsection{Baseline model performance}\label{baseline_performance}
Overall, the results show that the proposed model can learn well from the relatively small training dataset and produces physically sensible parameters and residuals.

\begin{figure}[h]
    \centering
    \includegraphics[width=0.9\linewidth]{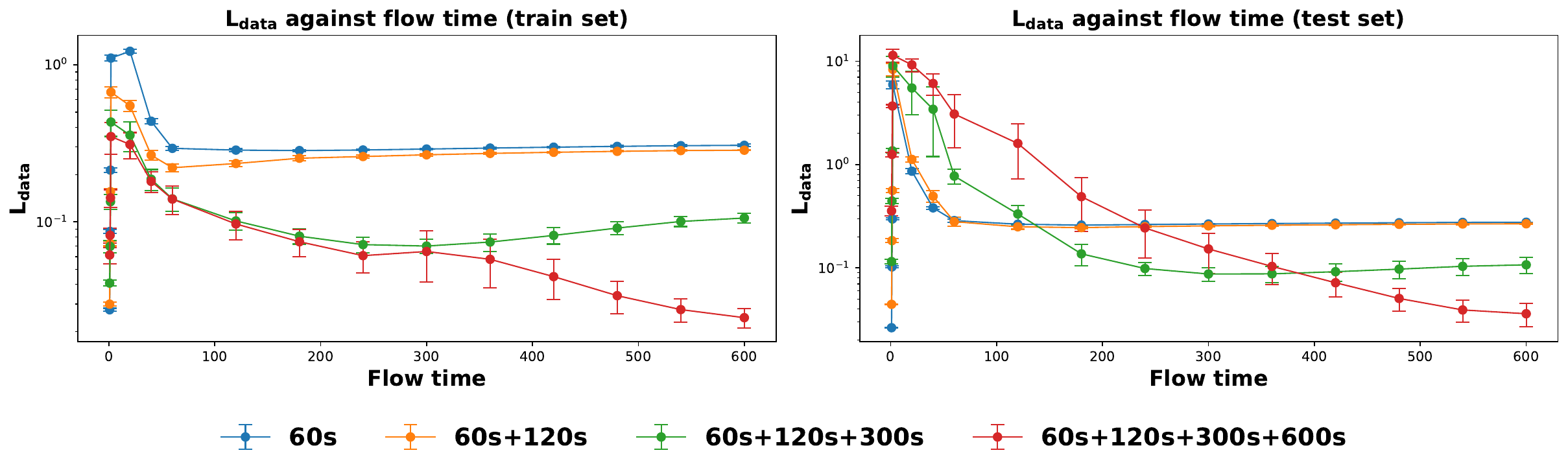}
    \includegraphics[width=0.9\linewidth]{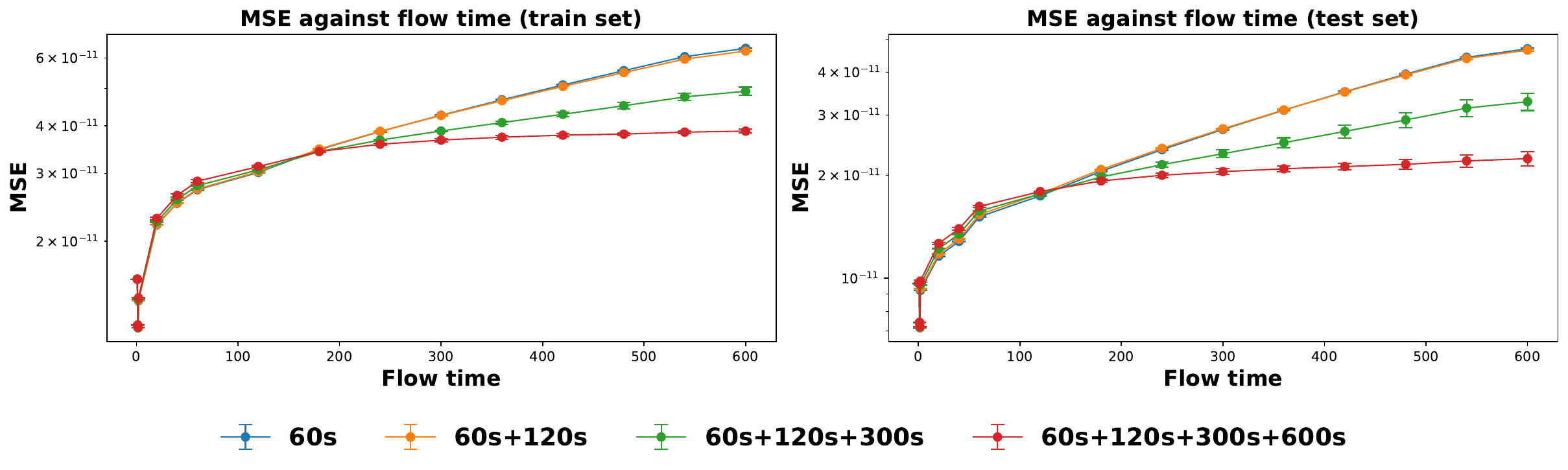}
    \caption{Comparison of $L_{data}$ and MSE against flow time for the (Left) training and (Right) test datasets, with different lines indicating results after each stage of curriculum learning (i.e. after being trained with 60s, 120s, 300s, 600s of flow time respectively). Loss is averaged over all different release locations of the training and test dataset respectively.}
    \label{fig:generalization}
\end{figure}

Curriculum training consistently reduces error for the longer rollout times, especially up to 600s, with improving generalization as evidenced by the decrease in test error in Fig.~\ref{fig:generalization}. The plots also illustrate relatively higher early-time relative error ($L_{data}$) due to the smaller denominators before the scalar spreads more widely. 

\begin{table}[h]
    \centering
 \caption{$L_{data}$ and MSE for the training and test datasets at 4 different simulation times (60s, 120s, 300s, 600s) after training}
\label{tab:baseline_performance}
    \begin{tabular}{>{\centering\arraybackslash}p{0.2\linewidth}cccc}\toprule
         &  60s&  120s&  300s& 600s\\\midrule
         $L_{data}$ (train set)&  $0.140$&  $0.097$&  $0.065$& $0.025$\\
 & $\pm0.029$& $\pm 0.020$& $\pm 0.023$&$\pm 0.003$\\
 \hline
         $L_{data}$ (test set)&  $3.084$&  $1.601$&  $0.153$& $0.036$\\
 & $\pm 1.639$& $\pm 0.872$& $\pm 0.064$&$\pm 0.009$\\
 \hline
         MSE (train set)&  $2.866\times 10^{-11}$&  $3.130\times 10^{-11} $&  $3.669\times 10^{-11} $& $3.872\times 10^{-11} $\\
 & $\pm 2.568\times 10^{-13}$& $\pm 1.789\times 10^{-13}$& $\pm 3.591\times 10^{-13}$&$\pm 4.486\times 10^{-13}$\\
 \hline
         MSE (test set)&  $1.621\times 10^{-11}$&  $1.791\times10^{-11} $&  $2.048\times 10^{-11} $& $2.235\times 10^{-11} $\\ 
 & $\pm 1.072\times 10^{-13}$& $\pm7.060\times 10^{-14}$& $\pm 3.944\times 10^{-13}$&$\pm 1.060\times 10^{-12}$\\ \bottomrule
    \end{tabular}
    
\end{table}

\begin{figure}[h]
    \centering
    \includegraphics[width=0.8\linewidth]{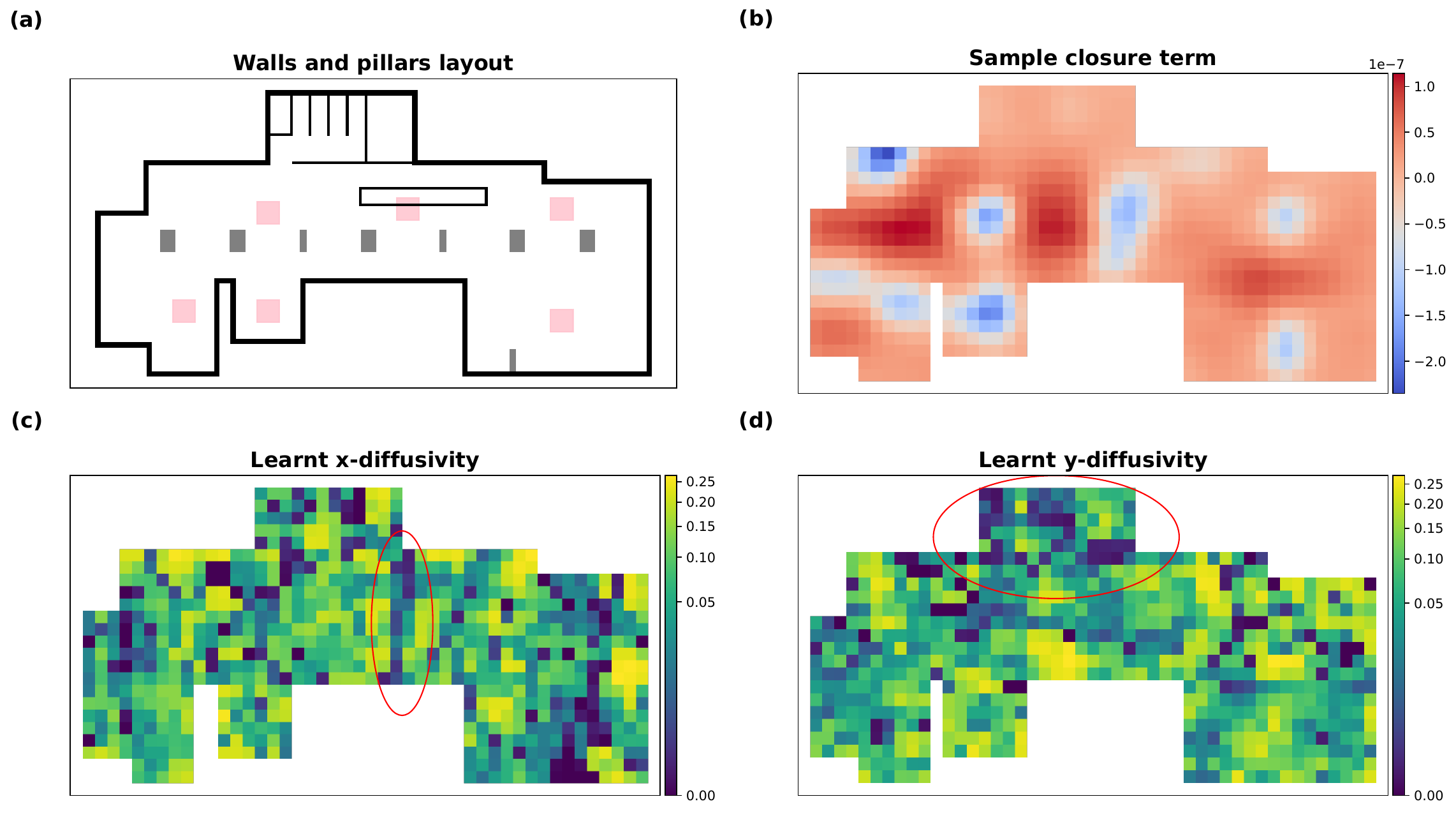}
    \caption{Learned diffusivity with some learned features circled in red and learned closure term. (a) Layout of room, with air-cons and pillars labeled as red and grey boxes respectively. (b) Sample closure term from a time point towards the end of the simulation time. Learned diffusivity along (c) x-direction and (d) y-direction. Yellow, green and blue indicate areas of high, medium and low diffusivity respectively. Low diffusivity regions (e.g. blue-black regions) can result if there are structural obstructions (e.g. internal pillars or walls).}
    \label{fig:qualitative}
\end{figure}

Qualitatively, the model learns diffusivity maps that line up with room structure as illustrated in Fig.~\ref{fig:qualitative}. We observe lower diffusivity near walls/pillars and an “air-curtain–like” region between air-conditioners and a small closure term where plane-normal transport due to the air-conditioning unit lowers the concentration, matching physical intuition. 

\FloatBarrier

\subsection{Physics-informed vs data-driven approach}\label{physics_informed_v_data_driven}
\begin{figure}[h]
    \centering
    \includegraphics[width=0.9\linewidth]{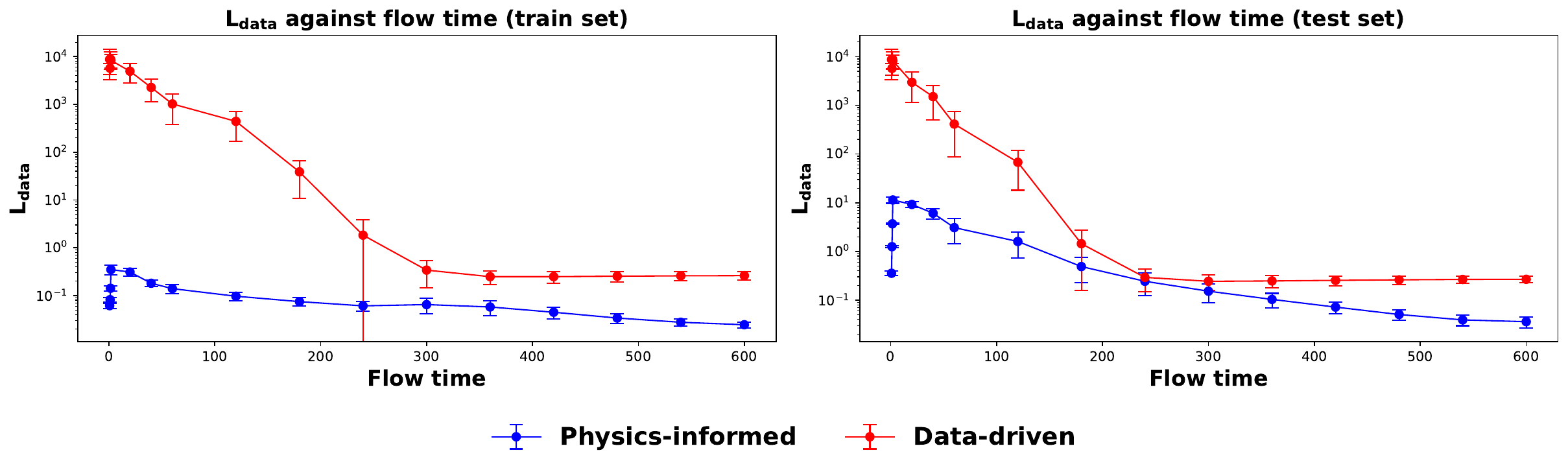}
    \includegraphics[width=0.9\linewidth]{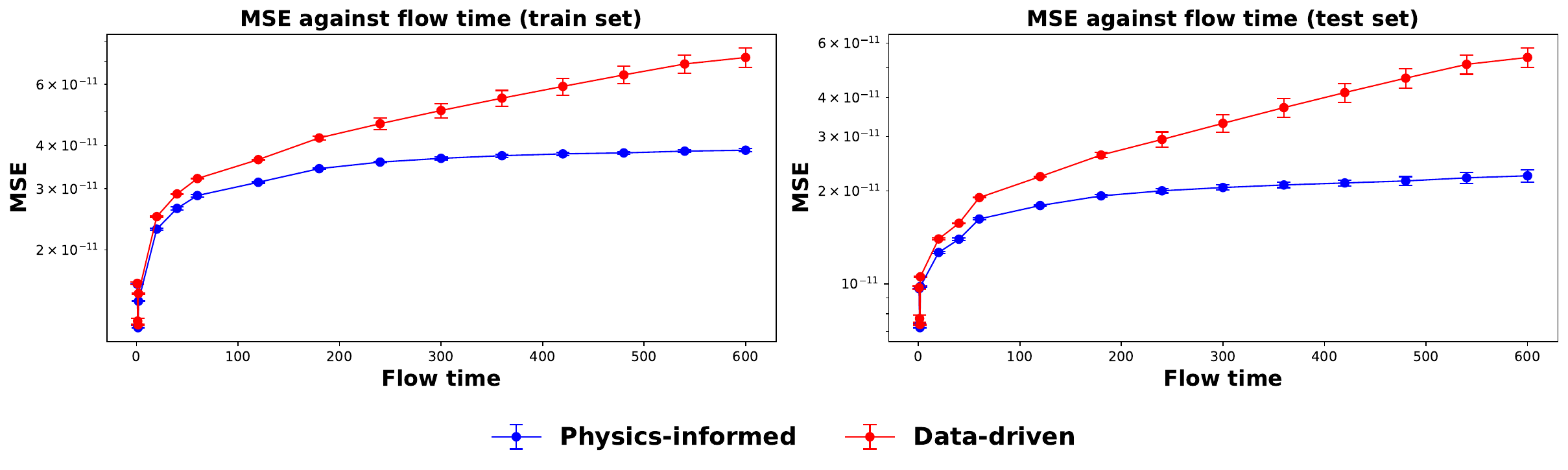}
    \caption{Comparison of $L_{data}$ and MSE against flow time for the (Left) training and (Right) test datasets, based on the physics-neural model and a solely data-driven approach. Errors are averaged over different release locations of the training and test datasets. }
    \label{fig:original_v_datadriven}
\end{figure}

\begin{figure}[h]
    \centering
    \includegraphics[width=0.9\linewidth]{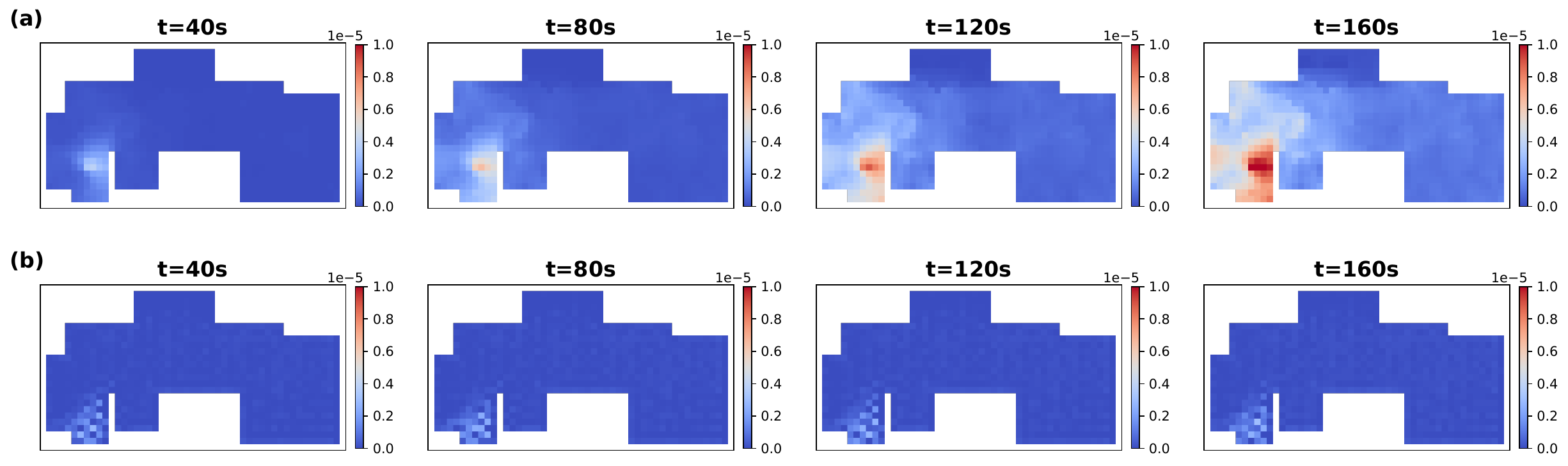}
    \caption{Comparison of simulation predictions on a sample test data using (a) the physics-informed approach and (b) the data-driven approach.}
    \label{fig:predict_original_v_datadriven}
\end{figure}

\begin{figure}[h]
    \centering
    \includegraphics[width=0.9\linewidth]{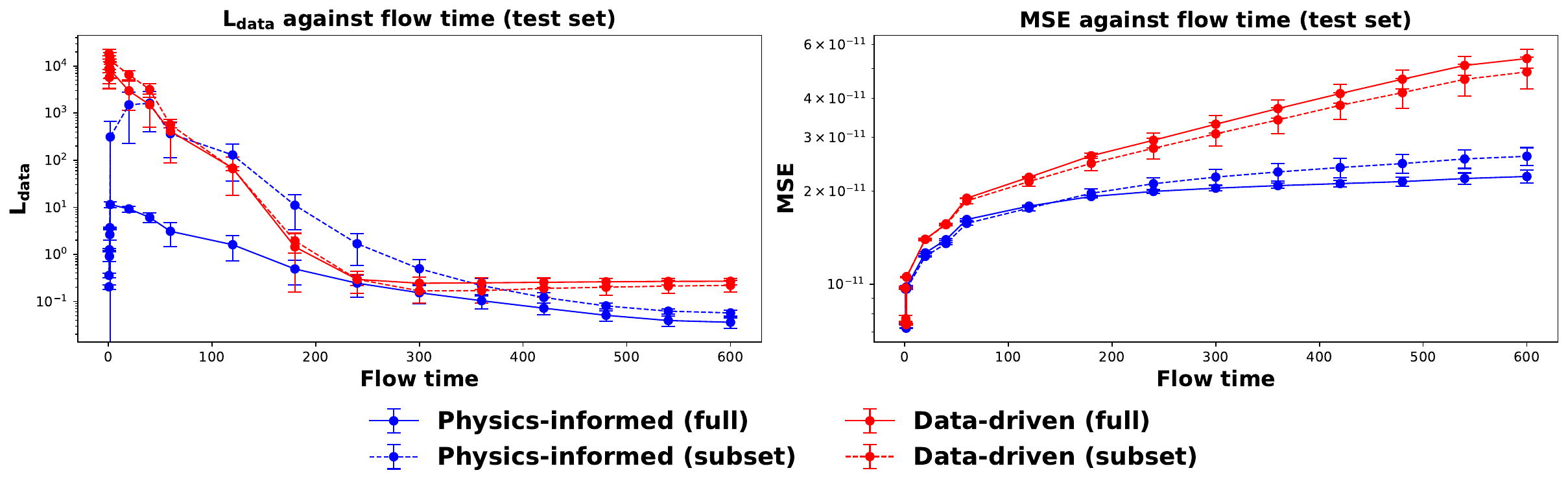}
    \caption{Comparison of $L_{data}$ and MSE against flow time on the test dataset when each model is trained using the full training dataset (26 simulations) or a smaller, randomly selected subset of the training dataset (8 simulations).}
    \label{fig:data_efficiency}
\end{figure}

We contrast the results from the full physics-neural model with a solely data-driven model in this section. The physics-neural model maintains low, steadily improving errors across long time horizons (Fig.~\ref{fig:original_v_datadriven} and yields physically consistent spatio-temporal fields as in Fig.~\ref{fig:predict_original_v_datadriven}. In contrast, the data-driven baseline shows large early-time relative errors and non-physical prediction fields. Even under further data scarcity (8 training cases), the physics-neural model exhibits improved prediction relative to the solely data-driven model as illustrated in Fig.~\ref{fig:data_efficiency}. 

After training, the MSE from the physics-neural model at the end of 600s is approximately half that of the data-driven model, as detailed in Table.~\ref{tab:original_v_datadriven}.
\begin{table}[h]
    \centering
\caption{Test $L_{data}$ and MSE for predictions using the physics-informed approach and data-driven approach at 4 different simulation times (60s, 120s, 300s, 600s) after training}
\label{tab:original_v_datadriven}
    \begin{tabular}{>{\centering\arraybackslash}p{0.2\linewidth}cccc}\toprule
         &  60s&  120s&  300s& 600s\\\midrule
         Test $L_{data}$ &  $3.084$&  $1.601$&  $0.153$& $0.036$\\
 (Physics-informed) & $\pm 1.639$& $\pm 0.872$& $\pm 0.064$&$\pm 0.009$\\
 \hline
         Test $L_{data}$ &  $413.118$&  $67.917$&  $0.243$& $0.268$\\
 (Data-driven) & $\pm 326.486$& $\pm 49.867$& $\pm 0.085$&$\pm 0.042$\\
 \hline
         Test MSE &  $1.621\times 10^{-11}$&  $1.791\times10^{-11}$&  $2.048\times 10^{-11} $& $2.235\times 10^{-11} $\\
 (Physics-informed) & $\pm 1.072\times 10^{-13}$& $7.060\times 10^{-14}$& $3.944\times 10^{-13}$&$1.060\times 10^{-12}$\\
 \hline
         Test MSE &  $1.900\times 10^{-11}$&  $2.222\times 10^{-11}$&  $3.302\times 10^{-11}$& $5.392\times 10^{-11}$\\ 
 (Data-driven) & $\pm 4.931\times 10^{-14}$& $\pm 8.402\times 10^{-14}$& $\pm 2.146\times 10^{-12}$&$\pm 3.834\times 10^{-12}$\\
 \bottomrule
    \end{tabular}
\end{table}

\FloatBarrier

\subsection{Non-Markovian vs Markovian approach}\label{nm_v_markov}
\begin{figure}[h]
    \centering
    \includegraphics[width=0.9\linewidth]{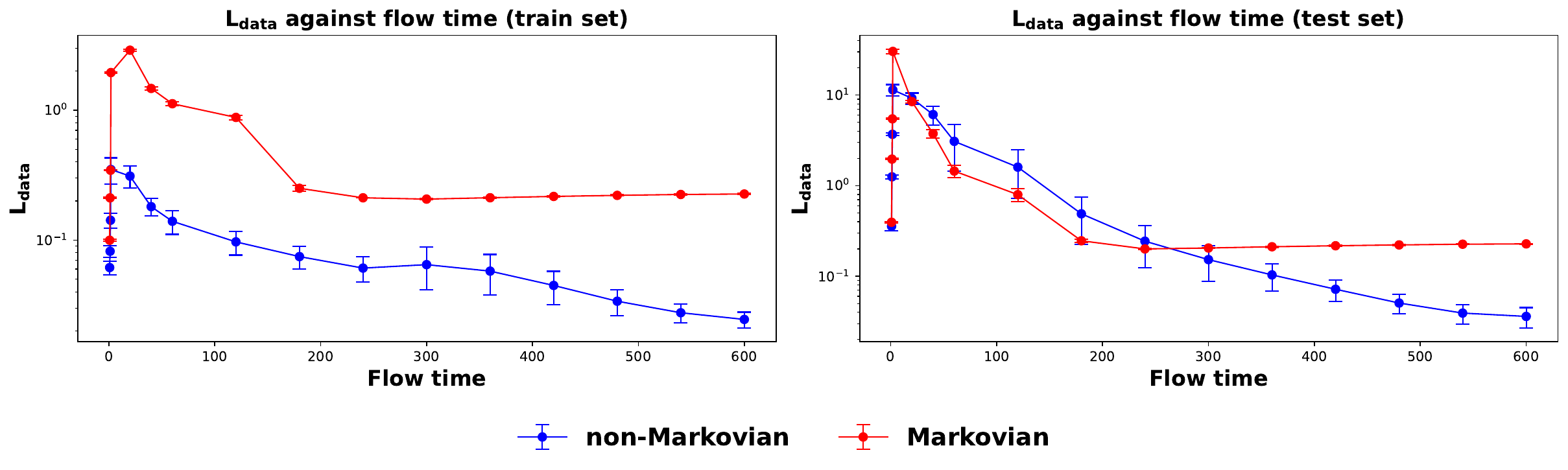}
    \includegraphics[width=0.9\linewidth]{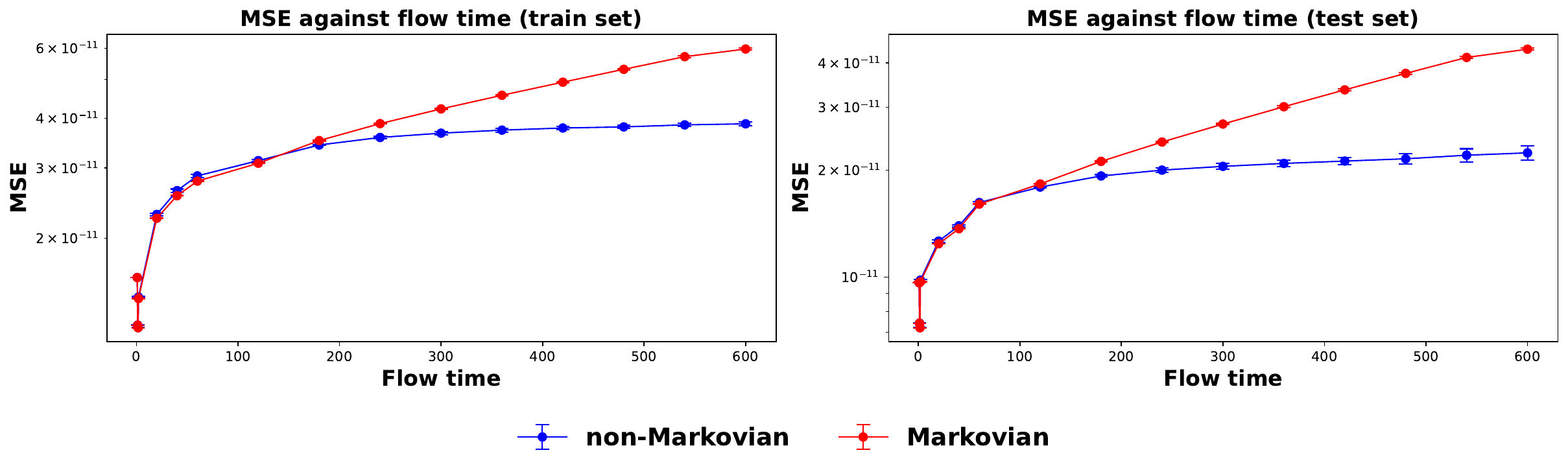}
    \caption{Comparison of $L_{data}$ and MSE against flow time for the (Left) training and (Right) test datasets, using the non-Markovian and Markovian models respectively. Metrics are averaged over different release locations in the training and test datasets.}
    \label{fig:original_v_markovian}
\end{figure}

\begin{figure}[h]
    \centering
    \includegraphics[width=0.9\linewidth]{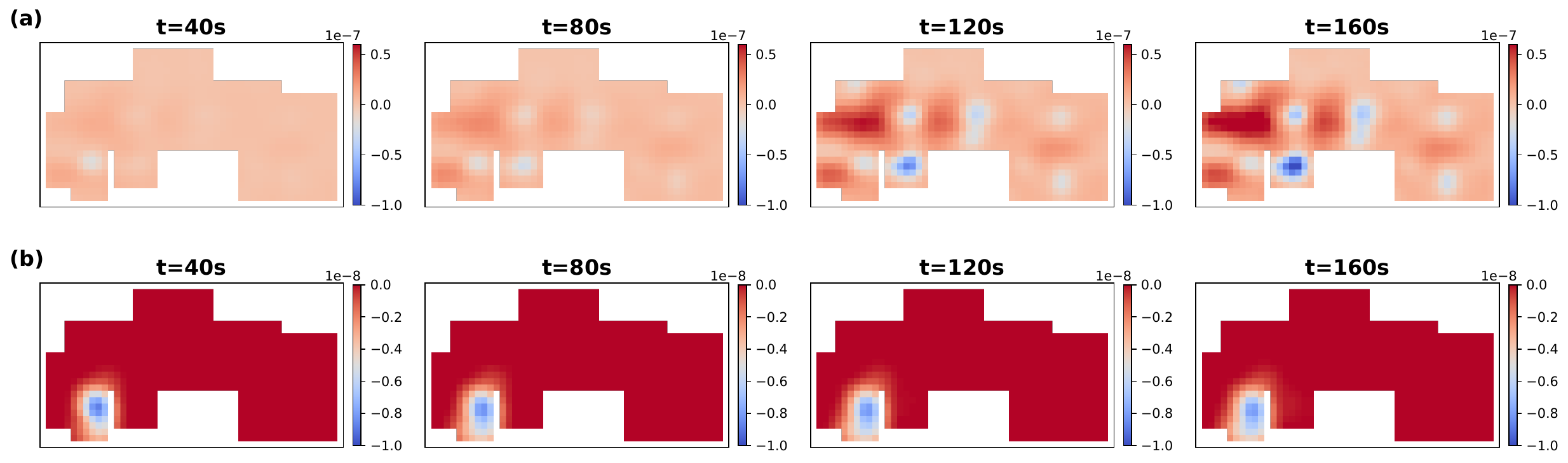}
    \caption{Comparison of predicted closure term on a sample test data using the (a) non-Markovian approach and (b) Markovian approach.}
    \label{fig:predict_original_v_markovian}
\end{figure}

Similarly, we compare the relative performance of the model when the history-aware component is removed from the model. In general, removing memory (i.e. a purely Markovian closure) yields competitive early-time losses but degrades at longer time horizons, where dependence on the historical trajectory is increasingly critical. The error curves separate as simulation time increases in Fig.~\ref{fig:original_v_markovian}, and closure visualizations in Fig.~\ref{fig:predict_original_v_markovian} show that the non-Markovian model produces coherent, trajectory-consistent closure patterns, while the Markovian equivalent does not exhibit any similar structure. This confirms the central claim as suggested by the Mori-Zwanzig formalism that memory is necessary to approximate unresolved effects, especially for longer time rollouts.

\FloatBarrier

Additional quantitative MSE metrics are provided in Table~\ref{tab:original_v_markovian}.

\begin{table}[h]
    \centering
\caption{Test $L_{data}$ and MSE for predictions using the non-Markovian and Markovian approach at 4 different simulation times (60s, 120s, 300s, 600s) after training}
\label{tab:original_v_markovian}
    \begin{tabular}{>{\centering\arraybackslash}p{0.2\linewidth}cccc}\toprule
         &  60s&  120s&  300s& 600s\\\midrule
         Test $L_{data}$ &  $3.084$&  $1.601$&  $0.153$& $0.036$\\
 (non-Markovian) & $\pm 1.639$& $\pm 0.872$& $\pm 0.064$&$\pm 0.009$\\
 \hline
         Test $L_{data}$ &  $1.449$&  $0.795$&  $0.205$& $0.228$\\
 (Markovian) & $\pm 0.231$& $\pm 0.125$& $\pm 0.003$&$\pm 0.002$\\
 \hline
         Test MSE &  $1.621\times 10^{-11}$&  $1.791\times10^{-11} $&  $2.048\times 10^{-11} $& $2.235\times 10^{-11} $\\
(non-Markovian) & $\pm 1.072\times 10^{-13}$& $\pm7.060\times 10^{-14}$& $\pm3.944\times 10^{-13}$&$\pm1.060\times 10^{-12}$\\
\hline
         Test MSE &  $1.604\times 10^{-11}$&  $1.827\times 10^{-11}$&  $2.693\times 10^{-11}$& $4.371\times 10^{-11}$\\ 
 (Markovian) & $\pm 3.221\times 10^{-14}$& $\pm 4.726\times 10^{-14}$& $\pm 1.253\times 10^{-13}$&$\pm 2.729\times 10^{-13}$\\
 \bottomrule
    \end{tabular}
\end{table}

\subsection{Joint learning of different physical parameterizations}\label{train_physics}
This section further isolates the benefit of jointly learning diffusivity and/or velocity parameters with the neural closure model. From Fig.~\ref{fig:physical_params}, we observe that learning neither greatly degrades performance. In contrast, the joint learning of both parameters yields the best improvement in long time horizon prediction. Joint learning of these parameters matter because they allow the model to implicitly encode for obstacles and flow behavior that were previously explicitly captured in the original CFD-derived quantities. Hence, allowing joint physics parameter learning can reduce the learning burden on the neural closure model, thereby improving stability and extrapolation.

\begin{figure}[h]
    \centering
    \includegraphics[width=0.9\linewidth]{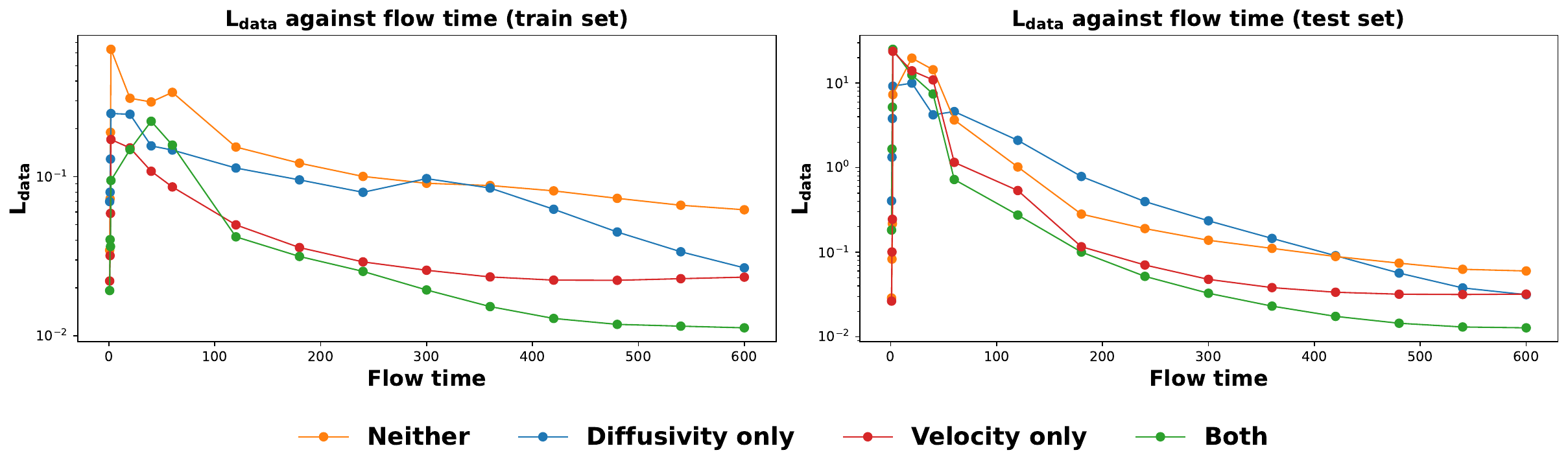}
    \includegraphics[width=0.9\linewidth]{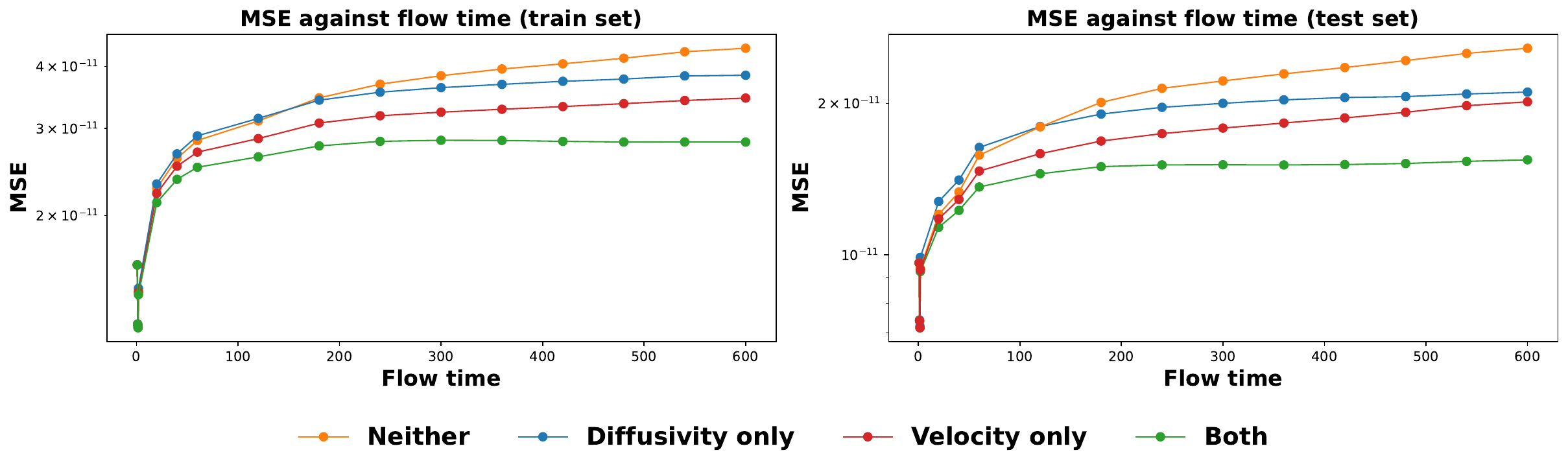}
    \caption{Comparison of $L_{data}$ and MSE against flow time for the (Left) training and (Right) test datasets, for models where different physical terms (diffusivity and/or velocity) are jointly learned with the non-Markovian neural closure model.}
    \label{fig:training_physics}
\end{figure}
\FloatBarrier

\subsection{Moving source scenario}\label{moving_case_scenario}
In this section, we investigate the ability of the trained model to generalize to new scenarios. In all previous training data simulations, the source was static and remained at the same location for all simulation time (0 to 600s). In contrast, this scenario involves a dynamically moving source. 

\begin{figure}[h]
    \centering
    \includegraphics[width=0.9\linewidth]{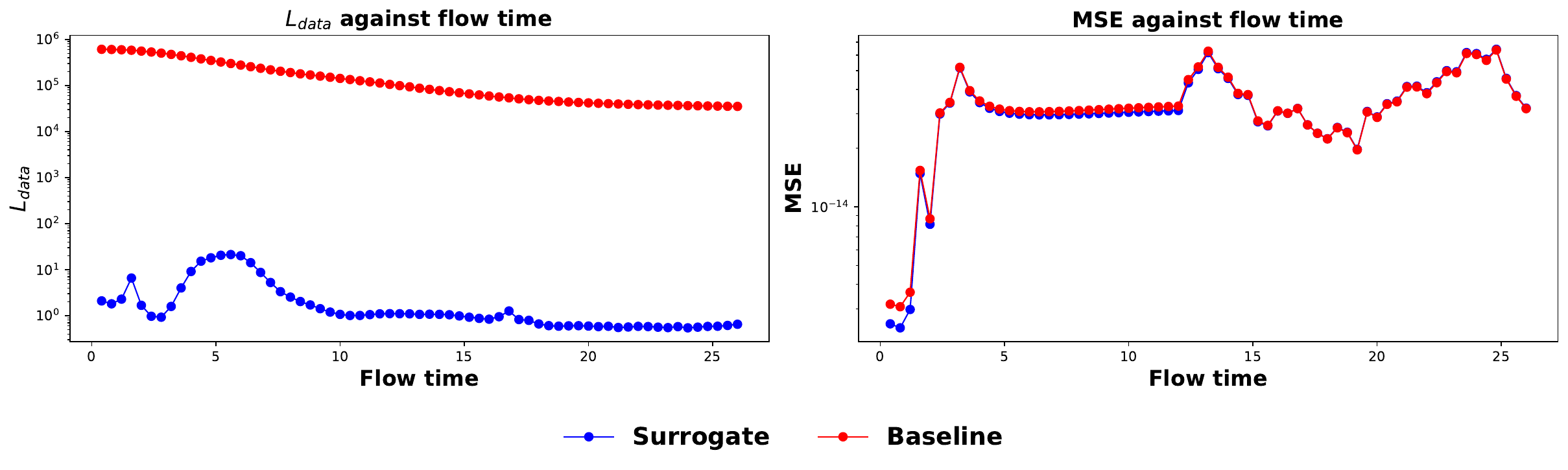}
    \caption{$L_{data}$ and MSE for a moving source scenario comparing predictions from the pre-trained surrogate model and a baseline where the concentration is assumed to be homogeneous and equal to the average concentration throughout the whole simulation.}
    \label{fig:transient_compare}
\end{figure}

\begin{figure}[h]
    \centering
    \includegraphics[width=0.9\linewidth]{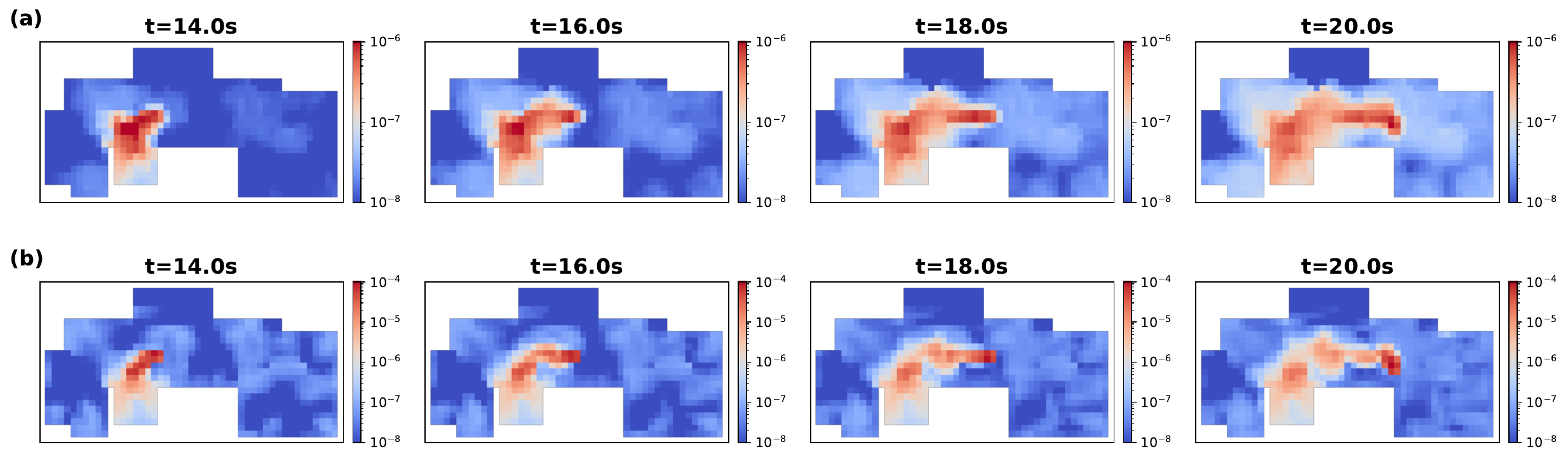}
    \caption{Snapshots of concentration field at different time steps (a) predicted by the physics-neural model (b) simulated by CFD. Note that the colors are on different scales.}
    \label{fig:transient_snapshots}
\end{figure}

\begin{figure}[h]
    \centering
    \includegraphics[width=0.6\linewidth]{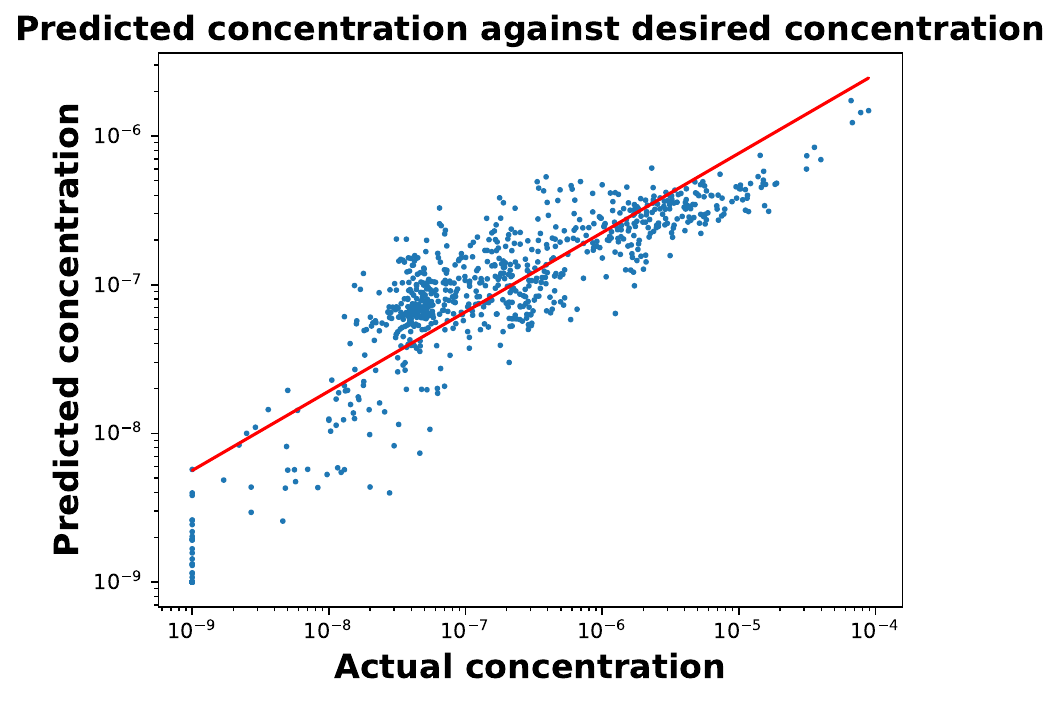}
    \caption{Comparison between concentration predicted by the surrogate model and the actual concentration at the final time step of the scenario of a moving source scenario. The Spearman correlation coefficient was calculated and found to be $0.957$. The red line shows the best linear fit on the log-log scale.}
    \label{fig:transient_correlation}
\end{figure}

Against a simple baseline, the pre-trained model can still achieve fairly low relative errors as shown in Fig.~\ref{fig:transient_compare}. The similarity in MSE is largely due to the difference in magnitudes between the prediction and the ground truth. Critically, the pre-trained model can reproduce salient field features as seen in the CFD snapshots in Fig.~\ref{fig:transient_snapshots}. The final-time Spearman correlation of $\approx 0.96$ in Fig.~\ref{fig:transient_correlation} further highlights the ability for the model to retain the right physical structure in the evolved field, although the predictive performance can still be further improved. This experiment suggests that this methodology can potentially yield more generalizable models while using a small amount of training data (26 simulations in this instance). 

\FloatBarrier

\end{document}